Running head: SIMDIALOG

SIMDIALOG: A VISUAL GAME DIALOG EDITOR

Charles B. Owen, Frank Biocca, Corey Bohil, Jason Conley

Michigan State University

East Lansing MI

Abstract

SimDialog is a visual editor for dialog in computer games.  This paper presents the design of SimDialog, illustrating how script writers and non-programmers can easily create dialog for video games with complex branching structures and dynamic response characteristics. The system creates dialog as a directed graph. This allows for play using the dialog with a state-based cause and effect system that controls selection of non-player character responses and can provide a basic scoring mechanism for games.

Introduction

A challenge in the design of computer games is writing, organizing, and testing non-linear dialog involving multiple user options and branches to many different possible character responses. One form of these are  dialog trees  (Bateman 2007) -- scripts that allow for multiple user inputs and character responses. We have developed SimDialog, a visual editor for complex non-linear game dialog.  The SimDialog system consists of an editor program that allows authors to design complex conversations with multiple options and a runtime component that manages the progression through the conversation during game play.  SimDialog makes it easy for game writers and non-programmers to create dialog for computer games directly without having to edit complex file formats or pass their work to programmers who must then embed the dialog in the game program.  It also provides a consolidated way to inventory all of the assets of a game specifically related to dialog, including audio clips and lip-sync data files.

Approaches to non-linear dialog in game design

A common theme in many video games is dialog between the player and non-player characters (NPC's).  Games such as *Leisure Suit Larry* (Lowe 2004) and *Desperate Housewives* (Games 2006) convey a story through the use of dialog.  Dialog may appear as text on the screen, audio recordings of spoken dialog, or fully lip-synced speaking characters.  As a participant in these conversations, the user impacts the story as presented, changing the way the story evolves or changing small details of the story.  Many games are dialog-intensive, with the majority of the player interaction taking place through dialog selection.

There are several ways that players can participate in dialog with a video game, the most basic are  menu selections and free text entry (Szilas and Kavakli 2006).  Allowing the user to type text directly into the system (or potentially speak to the system) permits great flexibility in the dialog.  However, this modality requires natural language processing and complex artificial intelligence concepts or the use of an extremely limited vocabulary and tolerance of considerable errors in understanding (Karlgren, Bretan et al. 1995).  For this reason, most games choose the alternative of presenting dialog options to the user in a menu, a method referred to as controlled interfaces.  It is this menu-based structure that is the subject of this paper.  The structure of a conversation then resembles a dialog graph with nodes representing utterances and links representing possible transitions.  A complete conversation is a path through the graph.

Authoring dialog for video games is a considerably more complex writing task than creating the script for a movie or television program.  The process of a conversation is nonlinear and cannot be captured as a simple script. Rather than a simple linear progression of topics, the game instead seeks to convey a set of scenes, themes, or ideas. Even if the NPC has only one path in the conversation, the player must have options from which to choose, so that the game creates the illusion of influencing story progression.  Hence, the dialog must be described in a more complex format than a script.

A common solution to this problem is the creation of a custom file format that supports the authoring process.  The SimDialog system is a game component that edits and organizes dialogue in a game.  It creates an XML file that contains all dialog for the game and can be loaded at runtime.  SimDialog assumes players and non-player characters (NPC).  Normally, only one player would be created, but that is not limited in the system.  SimDialog keeps a set of conversations that represent interactions between the player and each NPC.  Dialog is collected into pages for editing purposes only.

The structure of conversation is a directed graph.  No limits are imposed on the construction of this graph, though most games will rarely incorporate cyclic structures.

Most game engines, such as *Torque* and *Virtools*, embed dialog in scripts, requiring a programmer to translate the author's work into the game engine.  Dialog in video games is closely related in concept to interactive drama (ID), though the simple graphical structure common in computer games is typically deemed too limiting for modern ID.  The *PastMaster* system, for example, allows the creation of narratives, the equivalent of SimDialog conversations, using predicate logic (Szilas and Kavakli 2006).  Because all dialog is at the same level and no sample-once strategy is used, the number of text options in *PastMaster* grows rapidly, limiting the reasonable interaction period.

<div align="center">SimDialog Dialog Scripting Engine</div>

<div align="center">*SimDialog Elements*</div>

SimDialog collects *conversations*, interactions between a player and one or more non-player characters.  A conversation has a start and can have multiple ends.  The start may be invoked when the player approaches some NPC.  The end indicates the termination of the conversation, which may be an indication that a scene has ended or the NPC wishes to converse no more.  This section defines the elements of a SimDialog conversation.

*Player Character*. SimDialog assumes a player will interact with non-player characters. Decisions about what player line is to be spoken by the player character are made by the player via a menu on the screen.  Associated with a player are attributes that describe the player, such as the player name, gender, age, etc. SimDialog does not limit the number of players, though the system is best suited for single-player games.

*Non-player Character*. Non-player characters (NPCs) are the characters a player is interacting with.  NPCs have attributes similar to those of the player character.

*Actor.*  An actor is any character who can speak.  The set of actors is the union of player and non-player characters.  An Actor dialog box allows for the editing of actors and the indication of characteristics of the actor.

*Conversant.* A conversant is a non-player character that the player is interacting with. What a player chooses to say impacts the conversant, changing the conversant's state.  Normally, conversations will be among a player and a non-player character.  In this case the conversant is obvious.  However, conversations with more than two participants require some indication of to whom player statements are directed.

*Page.*  SimDialog organizes dialog into pages that can be used to organize scenes or any segment of dialog.  Multiple pages can be opened at once using the Window/New menu option. Each page is assigned a name.  Pages are only for organizational purposes, they do not restrict dialog functionality in any way.

*Dialog Modeling in SimDialog*

Dialog is organized in SimDialog in the form of graphs.  A graph starts with a named start node and can contain dialog items (cues), terminations, and references.  The items in the graph are called nodes.  There are several types of nodes.  Right-clicking the mouse on the workspace opens a context menu that includes options to create the available dialog nodes. Figure 1 is an example dialog graph. Double-clicking on a node invokes a dialog box that allows the node fields to be edited.

This game dialog structure if often referred to as a dialog tree (Bateman 2007). However, this is a misnomer when it applies to SimDialog, as a tree implies nodes can be reached from only one preceding node.  As can be seen in Figure 1, nodes (e.g., possible conversant responses) can be reached from multiple preceding nodes (e.g., a user's question).  A common criticism of a dialog tree structure is that it grows exponentially due to the branching.  In reality, the graph more commonly approximates a few parallel linear structures that navigate the player through the presented scene, perhaps including early exits due to player actions causing conversation terminations or even the player's demise.

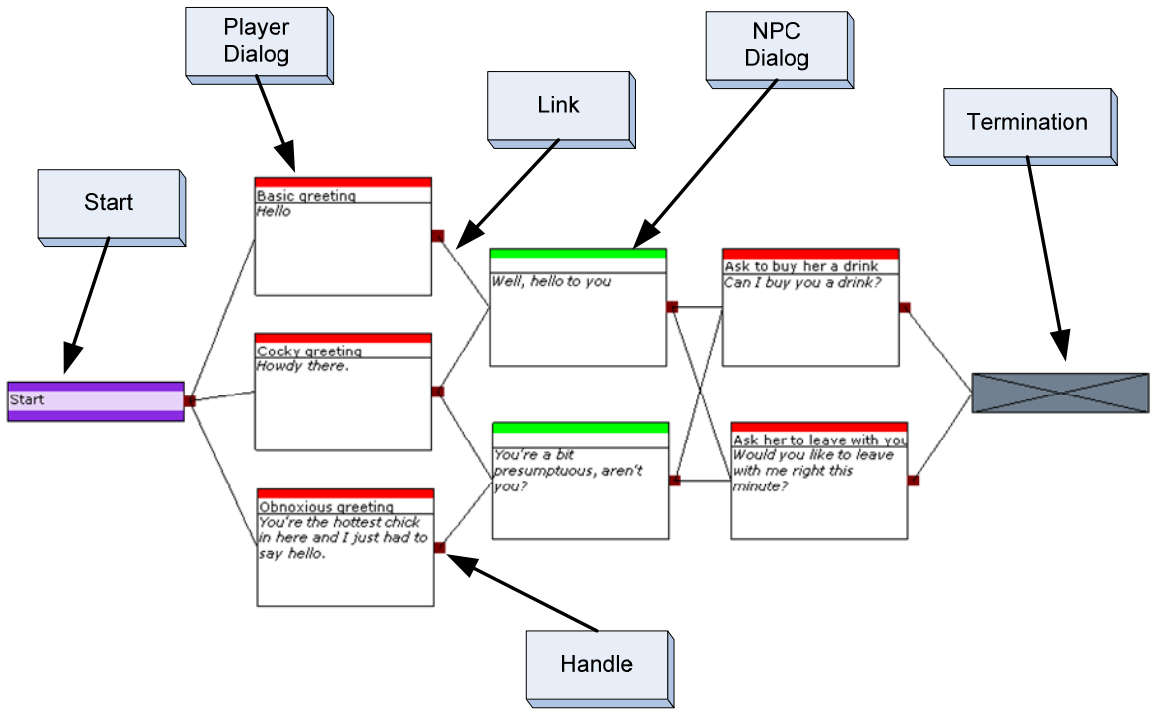

**Figure 1 - Example dialog graph**

*Dialog Item*

A dialog item is an element of speech that can be said; a cue. It is one thing that the player or NPC can say and includes any ancillary meta-data associated with the speech element. The colored bars in each dialog box indicate which actor is speaking the cue. The red bars in the figure represent *player* utterances, while the green bars represent a *NPC's* responses. Included with the cue is an optional *direction* field (not displayed in Figure 1), allowing for the inclusion of directions to the game engine. The direction field is analogous to a parenthetical in script writing. Parentheticals in stageplays describe stage directions (e.g., character moves to door, for example).  Either the direction or the cue can be omitted, allowing an NPC to response with shrug, a statement, or both.

Associated with a dialog item are the direction field, a label for the menu option, the actor and conversant, and names of associated files such as audio and lip-sync data files.

*Start*

A start node indicates an entry point into a dialog graph.  A name is associated with the start node and allows the game engine to choose an appropriate dialog at any given time.  Start nodes provide the game engine a named entry point for a conversation.

*Termination*

A termination node indicates the end of a dialog graph.  Termination nodes include information that tells the game that a dialog sequence is complete.  Termination nodes include a direction field that informs the game engine of the reasons for termination, so higher-level game logic can decide to proceed to a different conversation or scene.

*Reference*

A reference node is a link to another dialog graph.  It is an organizational tool designed to simplify the production of dialog graphs and to allow for transitions among scenes.  A reference node allows a longer conversation to be broken into pages.

*Subdialog Nodes*

Although the graphical structure of SimDialog allows a completely nonlinear dialog structure, conversations in video games are typically broken into sections that convey a specific idea or put the player to a specific test.  As an example, a conversation might consist of an initial meeting phase, a "getting comfortable phase", some social negotiation phase, and a terminal phase.  The initial meeting phase exchanges pleasantries such as names and basic information sufficient to determine if the conversation proceeds or terminates. The "getting comfortable phase" is the small talk that builds trust among the participants, the negotiation phase allows the participants to determine if the social arrangement will continue beyond this conversation and the terminal phase handles the parting. An author will typically conceive these phases independently, rather than as a complete conversation, particularly in a nonlinear environment such as a video game.

The subdialog node allows a unit of conversation to be treated as a single element in the dialog graph. The preceding example would be modeled as four subdialog nodes, each referring to a conversation on another page. The conversation proceeds from start to termination, and then continues to the next subdialog node. Termination nodes for a subdialog include a termination value, so that the subdialog node can be followed by multiple options, allowing the negotiation phase, for example, to proceed to a future date or to a slap in the face and immediate parting.

*Cause and Effect: How conversant states affect branching through the dialog trees*

A player will manually select from the multiple possible dialog options at any point in game play. However, the NPC selection must be made automatically. A greeting may be followed by a positive or negative response, depending on the "mood" of the NPC. SimDialog models this decision process through the use of states, causes, and effects. States are variables associates with actors that represent arbitrarily defined quantities such as mood, health, game score, etc. Causes determine how a specific state will be used to choose an NPC response. Effects are the impact upon the specific state that chosen responses illicit.

The states dialog box allows for the definition of states for the character and for non-player characters. States control the selection (i.e., probability) of which link will be used for an NPC (i.e., what the NPC will say or do). Player states are associated with the player. NPC states are associated with each NPC. There is one set of player states for the game and one set of NPC states for each NPC.  States are numbers that range from -1 to 1.  They will be displayed in the simulator (described below).

States are arbitrarily defined by the author and are specific to a game.  Example player states include "health" and "confidence."  Actions a player takes may increase or decrease health.  Responses from an NPC may increase or decrease confidence.  For example, an NPC saying "You look very nice tonight" might increase confidence; while a different NPC saying "You have no taste in clothes" might decrease confidence.  Example NPC states include "attraction" and "mood."  Attraction might indicate the NPC's attraction to the player's character.  Positive values indicate the NPC finds the player attractive, while negative values indicate the NPC finds the player repulsive.  The attraction an NPC has to the player's character may vary depending on what the player chooses to say.

States associated with an NPC may indicate states of the NPC or states of the character.  As an example, an NPC may have two attraction states -- one indicating the NPC's attraction to the player, the other indicating the player's attraction to the NPC.  SimDialog allows an arbitrary number of states to be defined for the NPC and the player.  State names are also arbitrary and determined entirely by the author.

Cause is associated with NPC dialog item and end nodes.  The selection of a player dialog element is performed by the player from a menu, so cause is not applicable and is deactivated in the item dialog box.  Cause is also not applicable to start nodes, since they are explicitly selected by the game engine to indicate a conversation start.  Effect is associated with start and dialog item nodes.  Selection of any dialog item causes an effect, be it selection by the player through a menu or by the system due to cause.  The effect associated with the start node allows an initial state to be defined.

*Cause*

A cause is how states will be evaluated in the process of choosing which dialog item to speak.  Causes are only used for NPCs.  There is a "General" cause weight and a cause weight for each player and NPC state.  Weights range from -1 to 1 and are indicated in the user interface by a slider control.  Figure 2 is an illustration of the dialog item dialog box.  Cause weights are on the left.

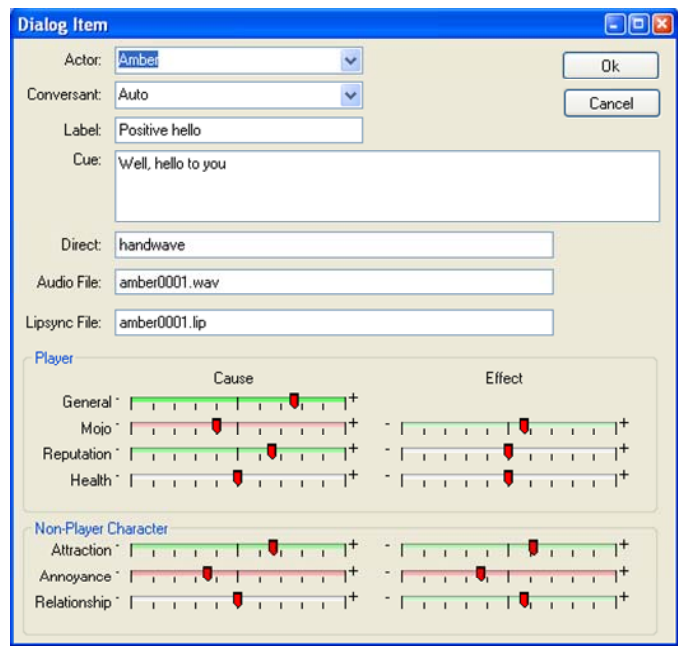

**Figure 2 - Dialog item dialog box (see text for details)**

After a player has selected a dialog item from the available choices, the weights for all following NPC dialog items are computed.  The weight is the sum of the general weight and the products of the current states and their associated weights (see Equation 1 below).

A positive weight indicates a given state will have a direct impact on the total weight.  A negative weight indicates a given state will have an inverse impact on the total weight.  For example, a positive weight for player confidence will indicate increased player confidence and increases the likelihood a given dialog item is selected.  This may be a dialog item that would be said to a player who is demonstrating confidence. A negative weight for confidence would indicate a dialog item that favors lack of confidence. Zero weights indicate a given state is ignored in the selection process.

A cause weight is in the range [-1, 1]. The dialog item has an associated set of cause weights $<w_g, w_p, w_n>$, where the scalar $w_g$ is the general cause, $w_p$ is a vector of cause weights associated with player states and $w_n$ is a vector of cause weights associated with the non-player character. Let $s_p$ be the vector of player states and $s_n$ the vector of non-player states for the current conversant. The score s for a dialog item is computed as the sum of the general cause weight and the inner products of the weights and the scores for the player and conversant:

$$s = w_g + w_p \cdot s_p + w_n \cdot s_n \qquad\qquad 1)$$

The impact of any state can be nullified by simply setting the associated weight to zero.

The average cause value indicates if the node response is positively or negatively impacted by the set of states. SimDialog uses this average to compute the background color of a node in the user interface. White is an indicator of neutral cause, shades of green represent positive causes and shades of red negative causes. As an example, the response to "How are you doing" might be "Very well, thank you" or "Get lost, creep" depending on the NPC mood state. A positive cause for mood will cause the former response (increasing mood increases the score), while a negative cause will illicit the latter (decreasing mood decreases the score). In the interface, these two options appear as green and red respectively.

*Effect*

An effect is how states will be changed following execution of a dialog item. A positive value increases the state. A negative value decreases it. Weights are bounded to range between - 1 to 1.  As an example, a positive effect weight for player happiness would indicate that this dialog item has increased the player's happiness.

An effect weight is a number in the range [-1, 1].  Let $e_p$ be the vector of effect weights associated with the player states and $e_n$ the vector of effect weights associated with the NPC. The new state is the sum of the current state vector and the effect vector bound in the range [-1, 1]:

$$s_i = \max(-1, \min(1, (s_i + e_i)))$$

2)

Inventory Generation

*Game Simulator*

Right clicking on any dialog node brings up a context menu for the node.  An option in that menu is "Play".  The play menu option invokes the game simulator.  This dialog box allows the conversation, starting at the selected node, to be played in a dialog-only mode.  This gives authors the ability to test the game play immediately without importing the result into the final game engine. During simulation, all state variables for the player and conversant are displayed and can be manually manipulated.  Figure 1illustrates the game simulator.

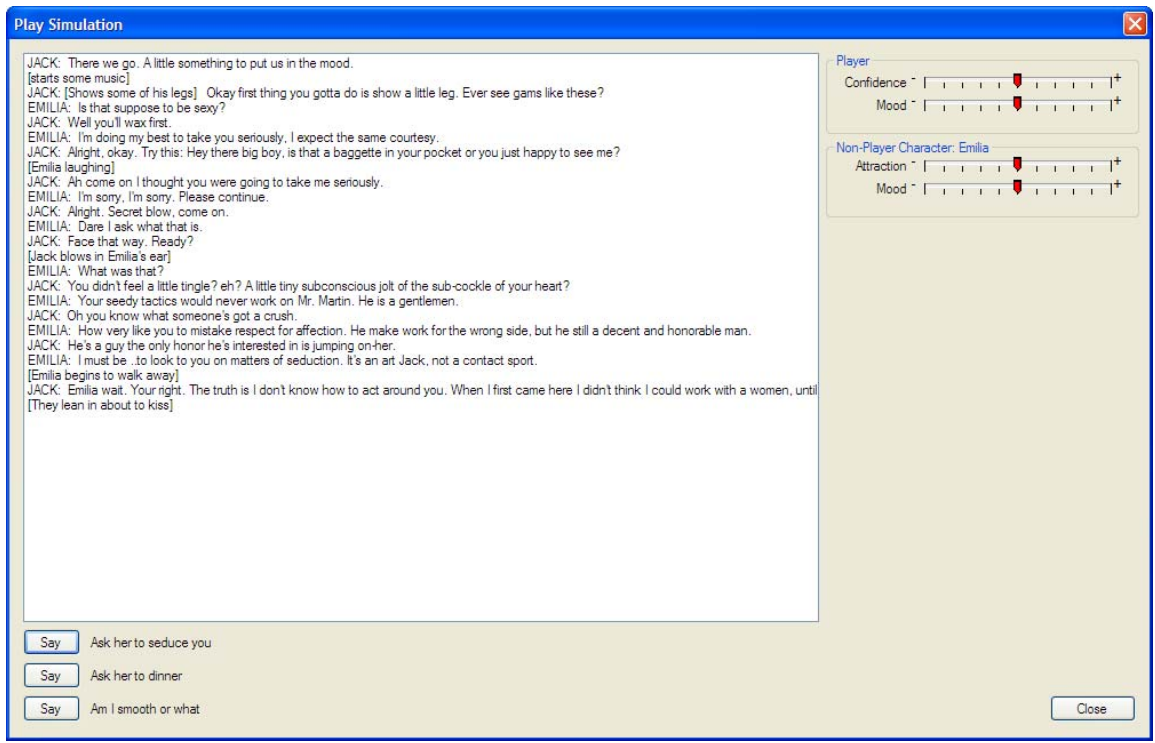

**Figure 3 - Game simulator**

Script Importing

Writing non-linear dialog can be quite complex.  Typically, a writer's initial draft of game dialog might be in the form of a linear script.  Authors create this form first to develop the general story line and flow of events.  In subsequent drafts different dialog paths are elaborated and appended. In traditional game development, this document is then annotated by the author to indicate branching and player options and then sent to programmers who subsequently transcribe the script into the form necessary for the game engine.

SimDialog allows authors to create scripts in a familiar form, and then *import* them into the system.  The importer reads a common script format, with actor-labeled lines and any parenthetical enclosed in brackets.  An initial game graph is created from the imported script. An example script and the imported graph are illustrated in Figure 4.

Jack: There we go. A little something to put us in the mood.

[starts some music]

Jack: [Shows some of his legs] Okay first thing you gotta do is show a little leg. Ever see gams like these?

Emilia: Is that supposed to be sexy?

Jack: Well you'll wax first.

Emilia: I'm doing my best to take you seriously, I expect the same courtesy. (Blake 2000)

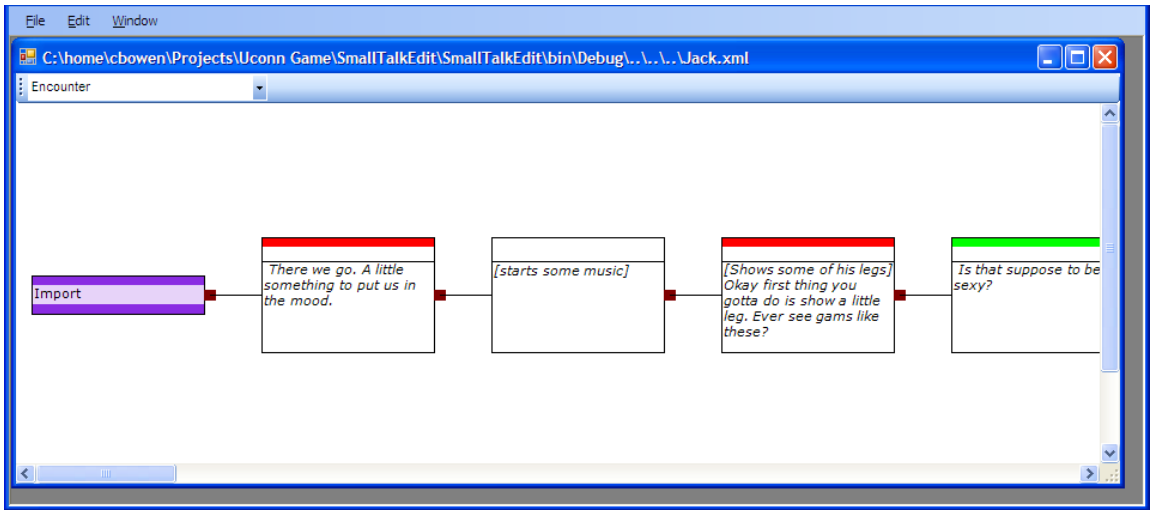

**Figure 4 - Example script and imported representation**

This feature then allows the writer to begin to more easily elaborate, draw, and visualize different non-linear paths through the conversation.  The cause and effect components of small talk allow the design team to define how each interaction affects player and NPC states.

## Conclusion

SimDialog is currently under development for a series of health communication games involving substantial interpersonal interaction.  This draft presents the state of the system and the current design goals.  The final submission of this paper will contain (1) additional narrative that discusses the writer experiences with the system, (2) integration with the Virtools game development system, and (3) demonstrations of the SimDialog dialog engine and simulator.

SimDialog is being created to solve a problem that is common in game development: how to enter large quantities of non-linear dialog so as to convey a story in conversations without requiring the direct participation of programmers.  As computer games become more widely accepted as tools for teaching and behavior influence, game development must become easier. Tools such as SimDialog and the experience gained in the program's development can help to drive development of tools that can help in that process.

Acknowledgements

This work was supported by grant #P01 CD000237 from the Centers for Disease Control.